\documentclass[fleqn,12pt,twoside]{article}
\usepackage[headings]{espcrc1}

\usepackage{graphicx}
\usepackage[figuresright]{rotating}

\newcommand{\AmS}{{\protect\the\textfont2
  A\kern-.1667em\lower.5ex\hbox{M}\kern-.125emS}}

\def\PLB{Phys. Lett. B}

\def\PRL{Phys. Rev. Lett.\ }
\def\PRD{Phys. Rev. D}
\def\PRC{Phys. Rev. C}

\def\etal{{\it et al.}}

\def\auau{$Au+Au$ }
\def\pp{$p+p$ }
\def\raa{$R_{AA}$ }
\def\sqrs{$\sqrt{s_{NN}}$ }
\def\pt{$p_{T}$ }

\title{PHENIX Results on Open Heavy Flavor Production and Flow in \auau Collisions at \sqrs = 200 GeV }

\author{Sergey A. Butsyk\address[LANL]{Los Alamos National
Laboratory, \\ MS H846 P-25 LANL, Los Alamos, NM, 87545, USA} for
the PHENIX Collaboration\thanks{for the full list of PHENIX authors
and acknowledgements, see Appendix 'Collaboration' of this
volume.}}

\runtitle{Open Heavy Flavor Production and Flow in \auau Collisions
at \sqrs = 200 GeV} \runauthor{S. Butsyk}

\begin{document}

% typeset front matter
\maketitle

\begin{abstract}
The paper presents recent PHENIX results for the nuclear
modification factor $R_{AA}$ in $Au+Au$ Collisions at $\sqrt{s_{NN}}
= 200$ GeV for the electrons originating from semi-leptonic decays
of open heavy flavor carrying particles. We also report the results
of the azimuthal asymmetry ($v_{2}$) measurement for those
electrons, which is directly related to the heavy quark elliptic
flow.
%The combination of those results is essential to understand
%the mechanisms of heavy quark energy loss in hot dense medium.
\end{abstract}

\section{Introduction}

Interaction of heavy quarks with hot dense matter, created in heavy
ion collisions at RHIC, is a very important probe for understanding
the properties of the produced medium. For the case of the light
quark, the strong suppression of high transverse momentum pions has
been experimentally measured \cite{ppg003,ppg014}. This important
result implies that hard scattered partons traveling through the
medium created in such collisions experience considerable energy
loss. The same effect for heavy quarks is predicted to be smaller
than for the light quarks due to a suppression of the phase space
for gluon radiation for large masses of the quarks (the so called
``dead cone'' effect \cite{dima}). Due to the interaction of heavy
quarks with the medium, heavy flavor particle is expected to exhibit
none zero elliptic flow $v_2$ \cite{flow}. The strength of the heavy
quark flow in combination with the degree of heavy quark suppression
provide crucial information on the nature of the heavy quark
interaction with the medium and are therefore very important for the
understanding of the medium properties \cite{derek}.

\section{Results on \raa}
 PHENIX has developed robust and
accurate methods \cite{ppg011,ppg035,ppg037} of measuring the heavy
quark production by disentangling the electron contribution from the
semi-leptonic decays of the open charm/bottom particles (this
contribution is denoted as \emph{``non-photonic''} electrons) from
the electrons created from the conventional sources (Dalitz decays
of light mesons, photon-conversions in the detector material -
referred to as \emph{``photonic''} electrons). The first method,
``cocktail'' subtraction, is based on simulating the ``photonic''
electron component with hadron decay generator using published pion
results \cite{ppg003,ppg014} as input. The second analysis
technique, ``converter'' subtraction, is based upon direct
measurement of the ``photonic'' electron contribution from a special
set of runs with increased conversion material budget.

 Here we present results obtained from the experimental data,
collected during the Year 2004 high statistics \auau PHENIX run and
Year 2003 \pp run. A minimum bias sample of $460\times 10^{6}$
events was analyzed for \auau using the Beam-Beam Counter (BBC) as a
trigger source.  Compared to the published \auau results
\cite{ppg035}, this gives a factor of 100 statistical improvement,
another factor of two due to increased acceptance, and the photonic
background was reduced by a factor of two because of the removal of
the central vertex detector. For the \pp results, the low \pt part
come from the $9.8\times10^6$ minimum bias events while the high \pt
part come from the events firing the level-1 electron trigger (ERT)
corresponding effectively to $2.3\times10^9$ minimum bias events.
Comparing to the previous results \cite{ppg037}, the luminosity is
increased five times and electron acceptance improved by a factor of
four. The total material budget of the PHENIX detector corresponded
to $X \approx$ 0.7\% $X_0$ radiation length from the beam pipe
($0.4\%$) and air ($0.3\%$).

\begin{figure}[ht]
\begin{minipage}[t]{0.45\linewidth}
\includegraphics[width=1.\linewidth]{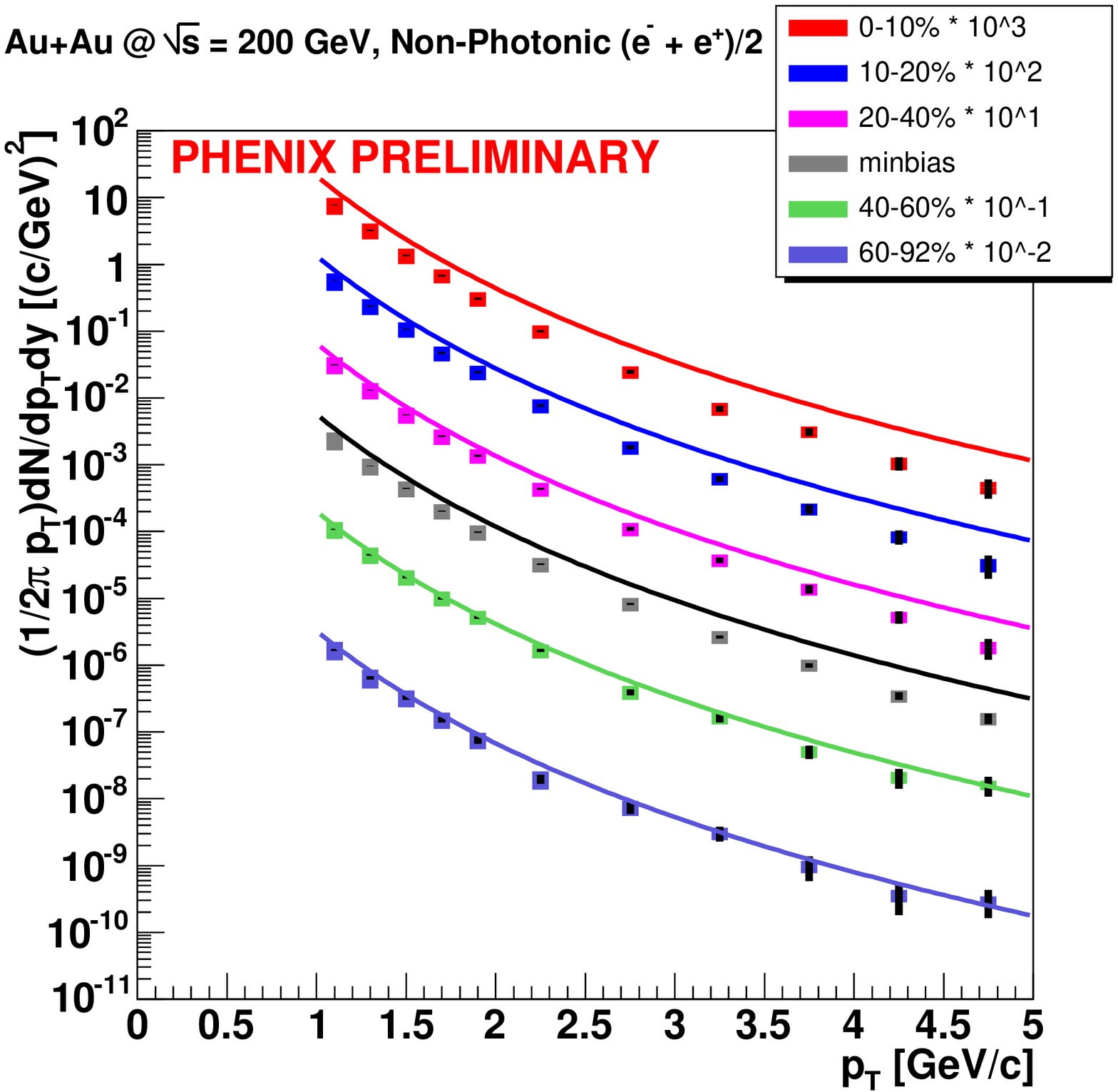}
\caption{``Non-photonic'' electron invariant yield in \auau for
different centrality bins compared with \pp best fit, scaled by
$N_{col}$.} \label{fig:non_photonic}
\end{minipage}
\hspace{\fill}
\begin{minipage}[t]{0.45\linewidth}
\includegraphics[width=1.\linewidth]{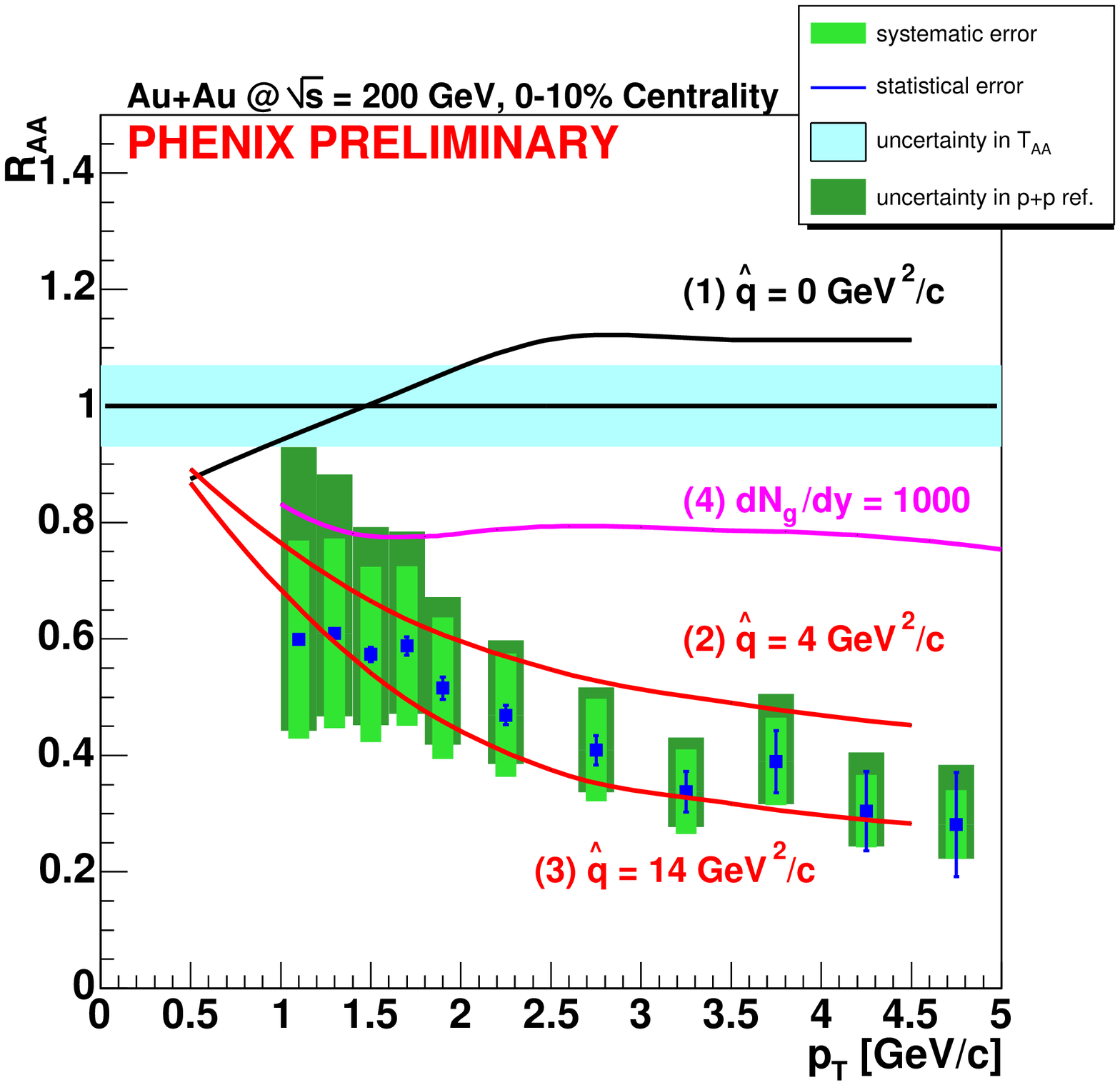}
\caption{``Non-photonic'' electron for 0-10\% centrality bin
compared with theoretical predictions. Theory curves (1-3) from
\cite{armesto}, (4) from \cite{magdalena}.} \label{fig:raa_0_10}
\end{minipage}
\end{figure}

The inclusive electron invariant cross section was corrected for the
acceptance and efficiency. The comparison of inclusive electron
spectra with the ``cocktail'' prediction for ``photonic'' electron
background clearly indicated an excess of the electrons. The
subtracted ``non-photonic'' electron yields for different centrality
bins are shown in Fig.~\ref{fig:non_photonic}. A fit to the \pp data
\cite{youngil}, scaled by corresponding number of binary collisions,
is also shown on the plot. One  can clearly see that for the most
peripheral bin ($60-92\%$) the \auau data agrees with the \pp
reference and that a high-\pt suppression of the electron yield
starts developing towards more central collisions.

To quantify the suppression, we use the nuclear modification factor
\raa \cite{ppg014} which is a ratio of the \auau invariant electron
yield to the \pp cross section, scaled by the nuclear thickness
function $T_{AA}$. \raa for the most central bin ($0-10\%$
centrality) is shown on Fig.~\ref{fig:raa_0_10}. The systematic
error for the ratio was separated in three pieces: the error on
$T_{AA}$, the systematic error on the data and on the \pp reference.
The comparison with the published theoretical predictions
\cite{armesto,magdalena} shows that the data supports theories with
large transport coefficient ($\hat{q} = 14 GeV^{2}/c$) or a very
high initial gluon density ($dN_g/dy = 3500$). This is a very
challenging task for a theory to justify such high values of these
parameters, but realistic values of gluon densities ($dN_g/dy =
1000$) significantly underestimate the heavy flavor suppression. It
is also very important to mention that in order to make a reasonable
theoretical prediction, one needs to include the electron
contribution from open charm and open bottom semi-leptonic decays.
%Due to its large mass, the bottom quark should experience smaller
%energy loss than the charm quark but the crossover point between
%bottom and charm contribution to the electron spectrum needs to be
%known with relatively good precision.

\section{Results on heavy flavor flow}

\begin{figure}[ht]
\begin{minipage}[t]{0.45\linewidth}
\includegraphics[width=1.\linewidth]{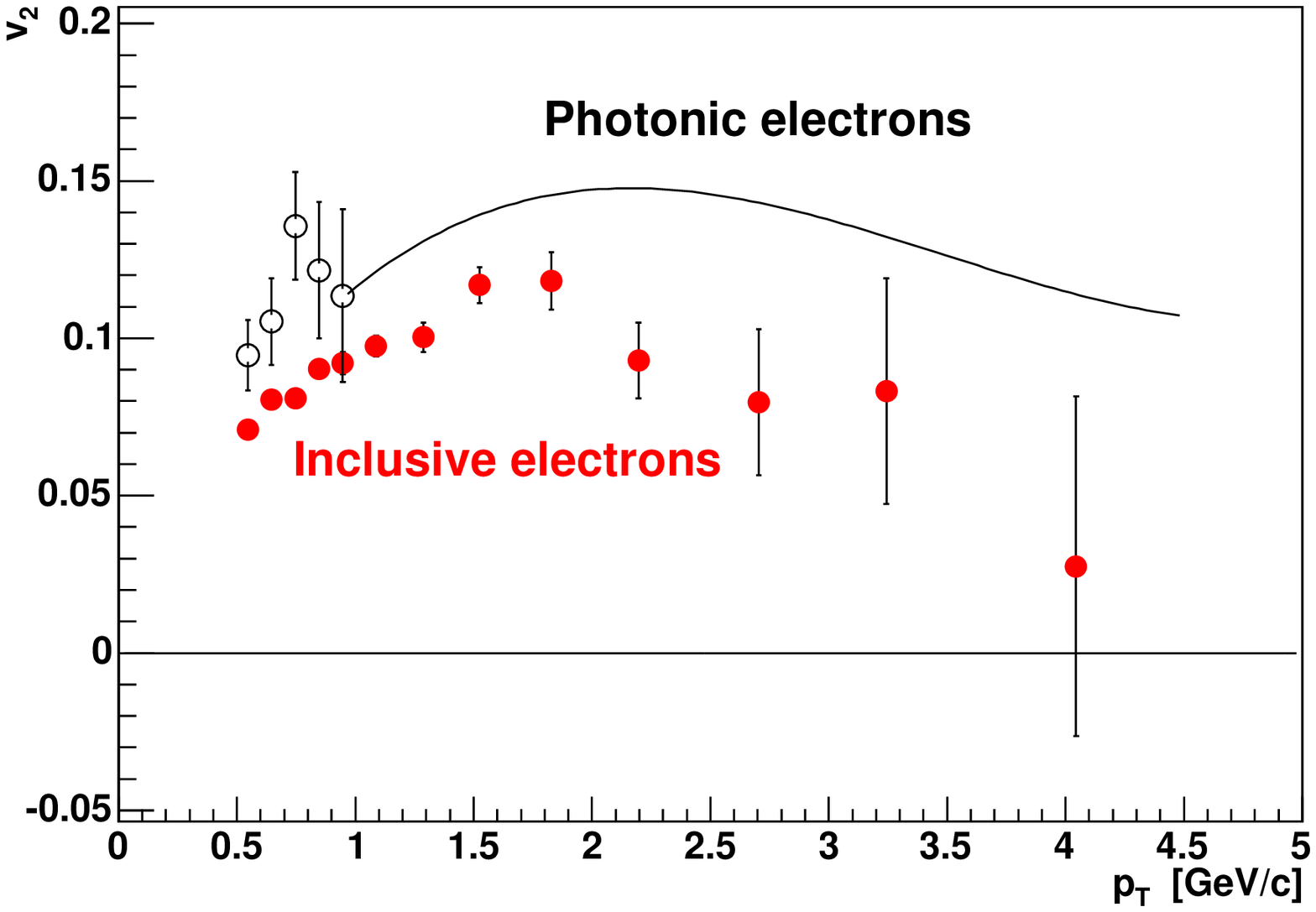}
\caption{Inclusive electron elliptic flow overlayed with
``photonic'' electron $v_2$ estimation.} \label{fig:flow_inclusive}
\end{minipage}
\hspace{\fill}
\begin{minipage}[t]{0.45\linewidth}
\includegraphics[width=1.\linewidth]{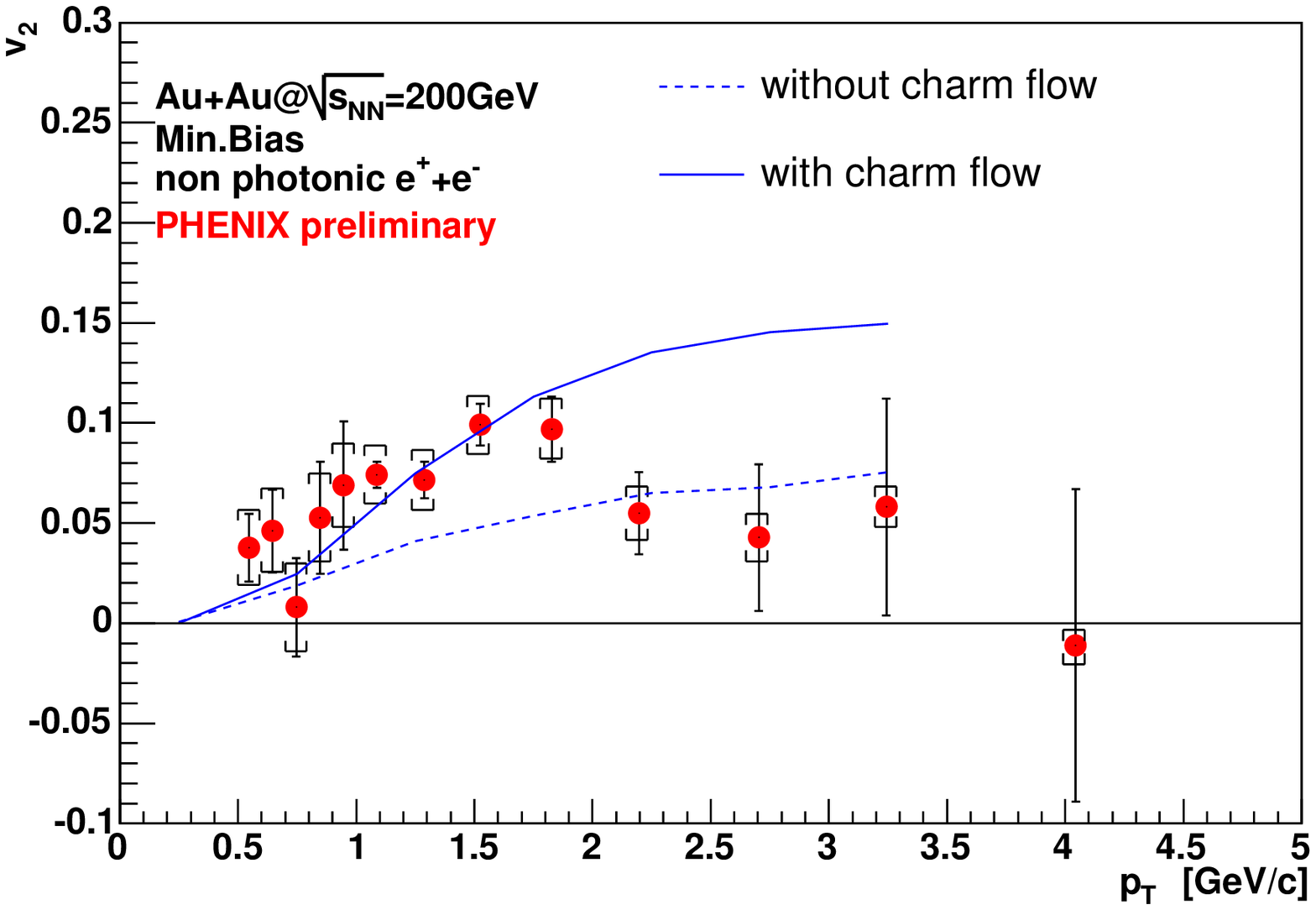}
\caption{``Non-photonic'' electron $v_2$ compared with theoretical
predictions for charm quark flow \cite{rapp}.}
\label{fig:flow_nonphot}
\end{minipage}
\end{figure}

Azimuthal flow of the open charm/bottom particles can be studied
indirectly by disentangling $v_{2}$ of the ``non-photonic''
electrons. Published results from the Year 2002 PHENIX run
\cite{ppg040} did not provide enough resolving power to make a clear
statement on whether the heavy quarks flow. Using the much higher
statistics of the Year 2004 \auau run, we obtain much more
significant results. The basic idea of the measurement is clearly
shown in Fig.~\ref{fig:flow_inclusive}. The inclusive electron
$v_{2}$ was measured from the data, then the ``photonic'' electron
$v_{2}$ was estimated using an analog of the ``converter''
subtraction technique for low \pt and simulated through the
``cocktail'' from measured flow of the pions for high \pt. Then the
flow of ``non-photonic'' electrons can be calculated, knowing the
relative contribution of ``non-photonic'' electrons in the inclusive
spectrum.

Results of this analysis are presented in
Fig.~\ref{fig:flow_nonphot} in comparison with theoretical
calculations for the electron $v_{2}$ under the assumptions that the
charm quark participates in the collective flow or just picks up the
flow from the light quark. One can see that at \pt $< 2$ GeV/c the
data are definitely in good agreement with the theoretical
predictions in which charm flow is assumed. However in the high \pt
region though, the magnitude of the flow saturates and decreases.
One possible explanation of this behavior is a contribution of
electrons from open bottom decay. The bottom quark is assumed to
have a lower $v_{2}$ which may cause the large \pt drop of the
``non-photonic'' electron elliptic flow.

\section{Summary and Outlook}

During the Year 2003/2004 run the PHENIX experiment accumulated a
significant amount of electron data in both \auau and \pp colliding
systems. New data dramatically improved the statistical significance
of the previously published heavy flavor measurements in the single
electron channel. Observation of strong suppression of the
"non-photonic" electron spectrum developing towards more central
collisions is a very important result, providing significant
constraint on the theoretical description of heavy-quark energy loss
mechanisms. The results of the "non-photonic" electron elliptic flow
measurements clearly indicate sizable $v_2$ of the open charm
particles.

\end{document}